# Shapes and velocity relaxation of dislocation avalanches in Au and Nb single crystals

G. Sparks[1] and R. Maaß[1a]

[1] *Department of Materials Science and Engineering and Frederick Seitz Materials Research Laboratory, University of Illinois at Urbana-Champaign, Urbana, IL 61801, USA*

We capture the spatiotemporal velocity dynamics of dislocation avalanches in face-centered cubic (FCC) gold and body-centered cubic (BCC) niobium crystals by compression testing of cylindrical microcrystals. In niobium, avalanche peak-velocities are over one order of magnitude lower, have distinctly rougher avalanche shapes, and relax more slowly to zero velocity than in gold. The avalanche dynamics (including averaged avalanche shapes) can be described reasonably well by mean-field predictions for avalanches near the depinning transition in the case of gold, but not for niobium. A detailed analysis shows consistent deviations (non-trivial exponents) from the predicted functional forms for both gold and niobium if the exponent that describes the velocity decay of the shape function is treated as a free parameter. While the avalanche size statistics and related scaling exponents are similar, these noticeable differences in the dislocation-avalanche dynamics demonstrate material-specific variations not predicted by "universal" behavior. The marked difference in avalanche shapes is discussed in terms of edge- and screw-dislocation mobility of an operating spiral arm source.

a) electronic mail: rmaass@illinois.edu

# 1. Introduction

Deformation of crystals is mediated by the motion of dislocations in response to an applied stress. At the macroscopic scale, the average velocity of a very large number of dislocations has long been used to motivate homogenization of plastic flow. This resulted in the formulation of constitutive laws that rely on large-scale averaged quantities with defined mean values. Deviations from Gaussian behavior are manifested in scale-free scaling of properties obtained from deformation experiments, as prominently shown with acoustic emission on hcp single crystals [1], or from a discontinuous stress-strain response in micro-straining experiments [2]. In contrast to classical homogenization schemes, the emerging view is now that plasticity is fundamentally intermittent and in many cases found to be scale-invariant [3]. This observation has led to numerous investigations focusing on the statistical properties of slip-size magnitudes, and in particular power-law exponents that describe the scale-free distributions [4-8]. The slip-size statistics in micro-compression testing have been shown to be relatively insensitive to crystal structure [4, 5] or the effect of test boundary conditions [9-11]. Scaling exponents of acoustic emission burst-energies [1] and slip-size magnitudes [4-7] in some experiments have been reported to be close to 1.5, which would theoretically allow the placement of plasticity in a universality class shared by many other intermittent processes, such as the flow of disordered metals, granular materials, earthquakes, or Barkhausen noise [12]. The ubiquitous power-law scaling might then motivate the use of simple analytical models that can capture the statistical behavior of the evolving system under different external drives, where precise details of the microstructural complexity become irrelevant. However, this contrasts somewhat with the behavior of crystals as described by classical dislocation theory and materials physics, in which microstructural specifics are key [13, 14]. In fact, early work reporting scale-free signatures of plastic flow predicted that new plasticity models based on scaling-laws are within reach [1, 15], but any novel description of plastic flow must include the materials-specific signature expressing that deformation depends on material composition and processing. Consequently, investigating scaling behavior in plasticity must eventually focus on differences between materials [13], which is one of the motivations for the work presented here. Many of the aforementioned studies have focused on slip-size magnitudes of stress-strain discontinuities that are due to a sudden collective dislocation rearrangement, also referred to as a dislocation avalanche. In comparison to slip-size magnitudes, it has remained a challenge to trace the dynamics of dislocation avalanches due to the severe confinement in space and the very short time scales associated with collective dislocation motion.

Here, we present time-resolved dislocation avalanche velocity profiles obtained during plastic flow of Au and Nb microcrystals; that is, we focus on the slip *dynamics*. We expand the methodology outlined in Refs. [16-18] to the precise time-resolved tracing of intermittent events, and find distinct differences in velocity profiles and velocity relaxation between the two crystal types. It is found that dislocation avalanches in Nb crystals are approximately 1.5 orders of magnitudes slower, up to 20 times longer in duration, and their averaged velocity profile exhibits a distinctly stretched relaxation behavior when compared to Au. This difference is robust against different testing modes (open and closed loop, force and displacement control). We attribute these differences to an exhaustion of mobile edge-dislocation components in the early stages of an avalanche in Nb, leading to an avalanche tail dominated by the low mobility of screw-dislocations. Furthermore, we compare the statistical signature and scaling of the avalanche dynamics in both crystal types with predictions from a mean-field theory (MFT) for avalanches near the failure stress (or "depinning transition") [19, 20], finding noticeably material-specific behavior contrary to the theoretical predictions in the case of variables related to slip dynamics.

## 2. Experimental details, analysis methods, and theory

### 2.1 *Experimental details*

The data presented here were obtained from cylindrical microcrystals (Au⟨001⟩ and Nb⟨110⟩) of nominally 2 µm diameter and 6 µm height prepared by annular focused ion beam milling from a bulk single crystal. This fabrication method results in slightly tapered cylinders with a taper angle between 1°-2°, usually around ~1.5° for these samples. The purity of the bulk Au crystal was 5N (99.999%), and the purity of the bulk Nb crystal was 4N (99.99%, excluding 180 ppm Ta), with interstitial impurities of 25 ppm C and 15 ppm O. Both crystals were oriented to produce multiple slip, which was confirmed by scanning electron microscopy imaging of slip traces on the crystal surfaces after testing as well as x-ray diffraction. All tests were performed at room temperature (~22 °C). The crystals were compressed using a Hysitron TI-950 Triboindenter equipped with a flat punch tip under displacement control, with an applied displacement rate of 60 nm/s (strain rate of $10^{-2}$ $s^{-1}$) and data acquisition rate (DAR) of 16 kHz for the Au samples and a displacement rate of 6 nm/s (strain rate of $10^{-3}$ $s^{-1}$) and a DAR of 3.2 kHz for the Nb samples. The difference in strain rate and DAR is necessary due to a combination of machine data acquisition constraints and the differing speed of dislocation avalanches in the two materials. All microcrystals were compressed to a total engineering strain of 20-25%, with a total of 14 Au and 12 Nb samples tested. The displacement-time profiles of the straining experiments were analyzed to extract individual dislocation avalanches (slip events) and calculate the velocity-time profiles during those events. The avalanche size was defined as the difference between the displacement at the beginning and the end of the event.

*2.2  Analysis methods*

Raw displacement-time data is obtained from the TI 950 Triboindenter. The raw data is smoothed by applying a lowpass filter using the Matlab commands fir1() and filter(). fir1(N,fc) uses a Hamming window to design a set of $N^{th}$-order finite impulse response (FIR) lowpass filter coefficients with a cutoff frequency of *fc*, where *fc* is a number between 0 to 1 representing a fraction of the Nyquist frequency. Thus, the absolute filter cutoff frequency *FC* is directly related to *fc* and the data acquisition rate (DAR) by $FC = fc\,(DAR/2)$. Applying the lowpass filter using the filter() function introduces both a filter delay and transient spikes at the beginning of the filtered data, so $N$ data points are removed from the beginning to remove the transient spikes, and the time vector is shifted by $N/2$ data points to compensate for the filter delay (e.g., if the raw depth and time vectors are each 1000 data points long and $N = 30$, then the smoothed displacement-time profile uses smoothed depth points 31-1000 and raw time points 16-985). Fig. 2 shows both raw and smoothed displacement-time profiles (as well as velocity-time profiles), demonstrating the good correspondence of the raw and smoothed data. For the gold data, the filter parameters used were $N = 30$ and $fc = 0.2$ with a DAR of 16 kHz. For the niobium data, the filter parameters used were $N = 30$ and $fc = 0.05$ with a DAR of 3.2 kHz.

The velocity-time profiles are calculated from the smoothed displacement-time profiles using the movingslope() function (available on the Mathworks File Exchange, created by John D'Errico). This function calculates derivatives at each point in a curve by doing a least-squares linear fit to data in a window of specified size centered on that point. For the data used in this paper, the window size was dynamically calculated for each event based on the initially estimated duration of the event. The window width was set to be equal to a number of data points equivalent to 1/8th the duration of the event, with a minimum of 2 data points and a user-chosen maximum (intended to retain finer details in very long-duration events). At a window size of 2 the velocity calculation is effectively a two-point difference method. The maximum window size used was 9 for the gold data (equal to a duration of 0.5 ms given the 16 kHz DAR) and 13 for the niobium data (equal to a duration of 3.75 ms given the 3.2 kHz DAR).

The event extents are based on the smoothed velocity profiles. An initial velocity threshold is defined to locate event candidates (contiguous regions above the threshold), and the event extents are then recalculated by starting at the peak velocity value in the candidate region and expanding the extents in either direction until the velocity value falls below a secondary threshold, defined here as 10% of the peak velocity value in the candidate region. Events were also manually inspected to discard certain spurious results, e.g. indenter-tip oscillations from loss of contact during tip retraction, which would occur at the end of the test or during the feedback-controlled return to the nominal displacement value following a particularly large slip event.

*2.3 Theory*

Our experimental data was tested against a MFT that relies on a depinning transition, where the system gradually develops towards a critical point. This model has shown good agreement with both crystal plasticity simulations and some deformation experiments [5, 9, 21], even though recent 2D dislocation dynamics simulations suggest that avalanches in crystal plasticity are scale-free at every applied stress [22, 23] (as is the case also in the mean field model for hardening [3,15]). Within the MFT framework, deformation is described by long-range coupled weak spots that slip in an avalanche-like fashion ("slip avalanches" or "dislocation avalanches") and events whose statistics scale similarly, regardless of the precise nature of the weak spots. The predicted probability distribution scaling of event sizes (i.e., total amounts of slip during individual slip events) has been found in some experiments on intermittently flowing microcrystals [2, 5, 7, 9, 19, 20, 23]. A sensitivity of the event-size scaling to experimental drive rate for a given material has been reported [24], and studies of acoustic emission amplitudes have found an even greater range of non-universal scaling exponents [25-27] depending on a variety of parameters. For avalanche event-sizes binned according to the stress at which they occurred, the probability density predicted by the used MFT model scales as

$$P(S,\tau) \propto S^{-\kappa} f\big(S(\tau_c - \tau)^{1/\sigma}\big) = S^{-1.5} f(S(\tau_c - \tau)^2), \qquad (1)$$

where $S$ is the event size, $\tau_c$ a critical failure stress in the material, $\tau$ is the stress for a given bin, and $\kappa$ and $\sigma$ are scaling exponents, with MFT predictions for their values used here [19, 20]. The scaling function $f(x) \propto e^{-Ax}$ for a non-universal constant A. For the statistics of events at all stresses (i.e. a stress-integrated distribution), the probability distribution for event sizes is predicted to scale as

$$P(S) \propto S^{-(\kappa+\sigma)} = S^{-2}, \qquad (2)$$

with the complementary cumulative distribution function (CCDF) for stress-integrated event sizes calculated to scale similarly as $C(S) \propto S^{-(\kappa+\sigma-1)} = S^{-1}$.

More recently, predictions for the dynamic behavior based on the same model have been presented [20, 28]. In particular, the avalanche peak velocity is predicted to scale with event size as $V_{peak} \propto S^{0.5}$. Similar scaling laws have also been developed for the *averaged* velocity-time profiles of similar sized events [20, 28-36], predicting that the peak velocity and event duration of the averaged profiles scale as $\langle V_{peak} \rangle \propto S^{0.5}$ and $\langle T_{event} \rangle \propto S^{0.5}$, where $S$ is the mean size of the averaged profiles in a given size bin. This would allow for the use of a factor of $S^{-0.5}$ to collapse the profiles from different size bins onto one another. The collapsed profiles are themselves described by a particular shape over time that is represented by

$$\langle v \rangle(t) = At e^{-Bt^2} \tag{3}$$

with $A$ and $B$ being non-universal fit parameters [28-31].

## 3. Results and discussion

### 3.1 Au and Nb stress-strain curves

Figure 1 shows a typical engineering stress-strain curve for both a Au and a Nb microcrystal. Strain hardening is minimal for both materials, with a slight increasing trend along the Nb stress-strain curve. Other curves (not shown) obtained for Au are practically similar to the one in Fig. 1, whereas additional curves (not shown) for Nb demonstrate larger variations including locally flat or negative slopes of the stress-strain response, with the overall behavior being approximately of zero slope. Local variations in the stress-strain behavior are a result of repeated unloading that occurs after larger slip events, which is caused by the feedback loop during displacement-controlled straining. The latter is unique for the used steering mode, and it is noted that force-controlled straining with the same device would not produce any locally nagative strain-hardening rates due to the continuously increasing applied force. Despite this steering-mode dependence of the overall flow response, it has been shown earlier that the scaling behavior of slip-size magnitudes from the same ⟨001⟩-orientated Au is insensitive to the loading mode [9]. The primary difference between the overall stress-strain response in Fig. 1 is thus caused by the stochastic occurance of very large slip events that lead to a machine-controlled unload and a relaxation of the dislocation network in the strained crystal.

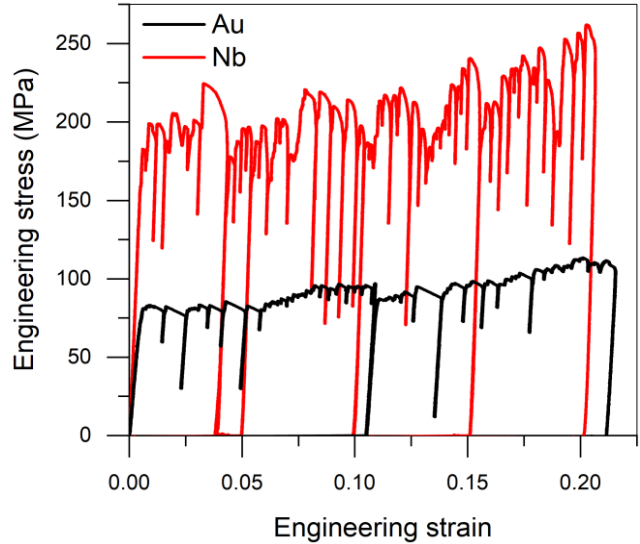

**Fig. 1.** Typical engineering stress – engineering strain curves for both gold and niobium.

*3.2 Au and Nb velocity profiles*

Typical velocity profiles from individual dislocation avalanches in both crystals are shown in Fig. 2. It can be seen that the peak velocity in the Au crystal is over an order of magnitude larger than that in the Nb, despite the similar size of the total net displacement (~40 nm) in both events. The Au velocity profile is smooth and fairly symmetrical during the slip event (Fig. 2a), whereas the corresponding velocity profile for the Nb crystal is slower and substantially longer in duration (Fig. 2b). This increased duration of slip events in Nb relative to Au (or other FCC materials) has been qualitatively observed earlier, but not investigated in more detail [37]. Furthermore, the avalanche velocity profile in Nb exhibits an increased relaxation time from its peak value and has a significantly greater degree of "roughness", i.e. the velocity varies more strongly throughout the event. Despite this variance, the avalanche velocity in Nb always remains much higher than the nominal displacement rate of 6 nm/s, consistent with the crystal staying in the avalanche state (crystallographic slip) throughout the event.

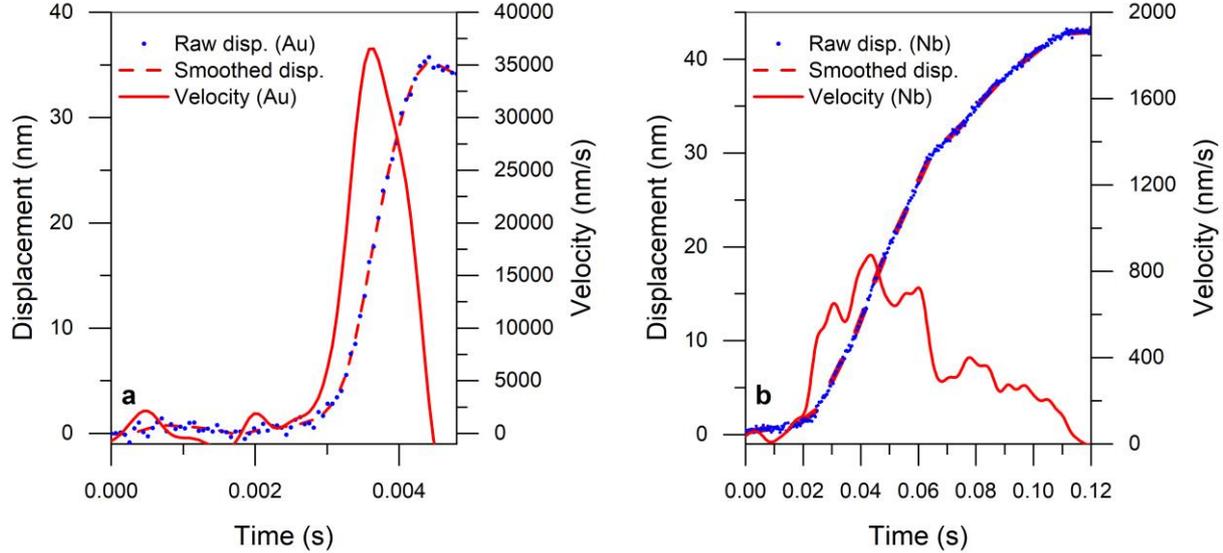

**Fig. 2.** Typical displacement and velocity profiles for individual dislocation avalanches in (a) gold and (b) niobium single crystal columns. Displacement and time values are shifted from raw data to begin at 0 nm/0 s for display. The Au event has a net displacement of 33.7 nm, and the Nb event has a net displacement of 41.9 nm. Both events are within the cutoff regions of the avalanche size distributions, well above their respective power-law scaling regimes (see Fig 3).

*3.3 Size distributions and related scaling exponents*

In order to demonstrate how the subsequent avalanche dynamics are evaluated, we begin with showing the complementary cumulative distribution functions (CCDFs) of event sizes for both crystal systems in Fig. 3. The range of event sizes is similar for both crystal systems, with a lower cut-off given by the minimum that can be distinguished from machine noise. The size CCDF scaling exponents were calculated using the Python `powerlaw` package implementation of the maximum-likelihood estimator [38]. This package also provides a statistical estimate of the error in the exponent, but this error is small relative to the changes that can be introduced by changing the limits of the fitted region [5] (see SOM for that paper), so the confidence intervals were instead calculated by manually varying both $X_{min}$ and $X_{max}$ and observing the range of exponents obtained. For both systems, the size CCDF is approximately linear on the log-log plot within a limited size range, demonstrating truncated power-law behavior, a result which has been theoretically predicted and experimentally observed previously [5, 7, 19, 34]. Table 1 shows the maximum-likelihood estimator analysis results demonstrating that a truncated power-law is the best match to the experimental data relative to other possible distributions via log-likelihood ratios and significance values [38]. The slopes in the linear regions are found to be $1.052 \pm 0.089$ (Au) and $1.077 \pm 0.116$ (Nb), which matches well with the scaling exponent predicted by MFT for a stress-integrated size CCDF, i.e. $\kappa+\sigma-1 = 1$. The cut-off regions at high event sizes that fall outside of the power-law scaling regime can be attributed to avalanche termination by finite size cut-offs, strain hardening in the material, and test system stiffness [7, 8].

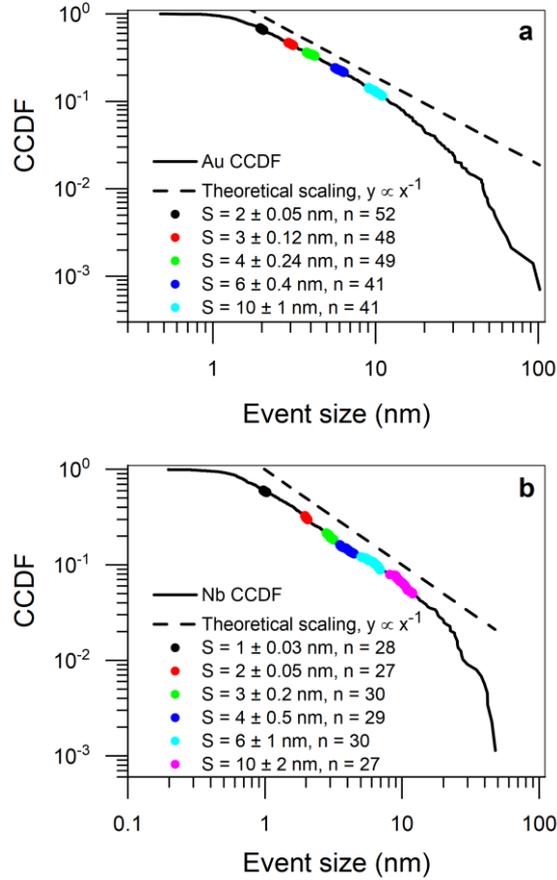

**Fig. 3.** Complementary cumulative distribution function (CCDF) power-law scaling of event size for events in (a) gold and (b) niobium microcrystals. The MFT-predicted scaling exponent of -1 is shown for comparison. The size bins used for the velocity profile collapse in Fig. 5 are highlighted, with the legend giving the bin ranges and number of events $n$ within each bin. See Table 1 for a maximum-likelihood estimator analysis of truncated power-law fits to the data in (a) and (b).

| Material | Distribution 1 | Distribution 2 | Log-likelihood ratio R (R > 1 favors Dist. 1) | Significance value p (p < 0.05 significant) |
|---|---|---|---|---|
| Au | Truncated power-law | Power-law | 5.006 | $1.093 \times 10^{-12}$ |
| Au | Truncated power-law | Exponential | 7.397 | $1.394 \times 10^{-13}$ |
| Au | Truncated power-law | Stretched exponential | 1.197 | 0.2312 |
| Au | Truncated power-law | Lognormal | 1.588 | 0.1122 |
| Nb | Truncated power-law | Power-law | 2.691 | $2.154 \times 10^{-3}$ |
| Nb | Truncated power-law | Exponential | 7.005 | $2.469 \times 10^{-12}$ |
| Nb | Truncated power-law | Stretched exponential | 4.234 | $2.296 \times 10^{-5}$ |
| Nb | Truncated power-law | Lognormal | 3.937 | $8.243 \times 10^{-5}$ |

Table 1. Maximum-likelihood estimation results for various hypothetical distributions for the data in Fig. 2.

## *3.4 Peak event velocities and size-velocity scaling*

The event size / peak velocity scaling shown in Fig. 4 also reveals large differences in the absolute size of the peak event velocities between the two systems. The difference in peak velocity between the two crystal systems is distinct, with the ratio between the peak velocities for events of similar size ranging between 30 to 50. Note that the exact values of the slip velocity are expected to be sensitive to the details of the analysis method that is used or the physical characteristics of the test setup, while the scaling behavior is predicted to be independent of such details. To evaluate the size-velocity scaling exponents, the raw size-velocity data was logarithmically binned in size, and the mean and standard deviation of size and velocity within each bin were calculated. A linear fit to the logarithm of the binned mean size and velocity values provided the $R^2$ goodness-of-fit values. To calculate confidence intervals for the scaling exponent, it was assumed that the measured variables were normally distributed within each bin. This allowed the generation of simulated data sets, with the size and velocity values for each bin pulled from a normal distribution with the same mean and standard deviation as the raw data for that bin. 10,000 data sets were generated and fitted with a linear fit to the logarithm of the generated data, producing an approximately normal distribution of exponent values. The confidence interval for the size-velocity scaling exponent was taken as the mean of this distribution ± 1.96 times the standard deviation of this distribution (i.e., covering 95% of the observed values). Neither fit precisely matches the predicted scaling of $V_{peak} \propto S^{0.5}$, with an experimentally observed power-law exponent of $0.656 \pm 0.076$ for the Au data (1424 data points, $R^2 = 0.99$ for bin-averaged data) and $0.437 \pm 0.071$ for the Nb data (879 data points, $R^2 = 0.97$ for bin-averaged data). For the Au data the predicted scaling exponent is entirely outside the experimental confidence interval, while for the Nb data it lies very near the limit of the confidence interval. In Fig. 4 we choose to present the as-obtained data for each individual avalanche so as to demonstrate the variability between different avalanches, even those of similar size in the same material.

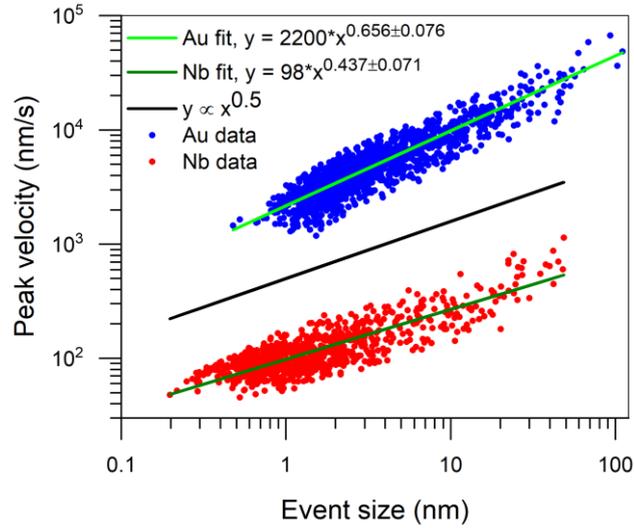

**Fig. 4.** Power-law scaling of peak velocity with event size for gold and niobium. The universal power-law exponent predicted by MFT is 0.5 and is shown by the line located between the data sets. Deviations from the MFT prediction of 0.5 are found for both crystal systems.

*3.5  Size-binned mean velocity profile shapes and velocity relaxation*

For further investigation of the differences in dynamic behavior between the two systems, we turn to an examination of the mean velocity profiles during avalanches, which according to Eq. 3 are predicted to have a specific shape and scaling behavior. To obtain mean velocity profiles for events of similar size, events are first binned according to event size (plastic displacement during the event). The velocity-time profiles of events within a given bin are padded with zero velocity values at the end until they are all the same length, shifted in time to begin at 0 seconds, and then directly averaged together to produce the mean velocity-time profiles. Error bars at each point are 95% confidence intervals based on the standard error of the mean for that point, i.e. $\pm 1.96\sigma/\sqrt{n}$, where $\sigma$ is the sample standard deviation and $n$ is the number of samples in the data used to calculate the mean and standard deviation. To obtain the parameters $A$ and $B$ for the nonlinear curve fits, the scaled curves for each size bin are combined into one data set and fitted using Matlab's fit() nonlinear model fitting function with the equations specified in Fig. 5. This function also provides the $R^2$ goodness-of-fit value. As shown in Figs. 5a and 5b, the collapse of profiles averaged across narrow bins centered around various event sizes S using the predicted scaling factor of $S^{-0.5}$ is good for both the Au and Nb data. Despite the greater variability in the Nb velocity profiles, in both crystal systems each of the mean profiles are within the confidence intervals of the other collapsed profiles for that system. In addition to the mean velocity profiles, the confidence intervals for each bin – themselves scaled by the factor of $S^{-0.5}$ – also overlap each other very closely. We note that the collapse works remarkably well despite the fact that the event size-velocity scaling seen in Fig. 4 does not precisely follow the predicted scaling exponent.

The predicted velocity profile shape-scaling function (Eq. 3) matches the velocity profiles from the Au crystals with a fitting quality of $R^2 = 0.98$. However, we find that the shape function describing the averaged/collapsed velocity profiles of the Nb crystals is distinctly different to the shape function for the Au data. The velocity profiles from the Nb crystals decay more slowly with time and are not as well fit by Eq. 3 ($R^2 = 0.84$), with even the best fit decaying more quickly than is observed, so a modified equation was tested. A much better fit ($R^2 = 0.95$) is found when using the form

$$\langle v \rangle(t) = Ate^{-Bt}, \tag{4}$$

which is similar to one previously used for average event shapes of magnetic domain avalanches [39, 40]. This function shows that even though both crystal systems have a similar scaling exponent for their event-size distribution, and also similar velocity profile behavior at low values of $t$, the dislocation avalanches in the Nb crystals relax considerably more slowly from their peak velocity than in Au crystals. This change in the velocity relaxation behavior was also found to be robust with respect to changes in the device control mode, including closed-loop displacement control, closed-loop force control, and open-loop force control, as long as the drive rate (in nm/s or µN/s) was not excessively high.

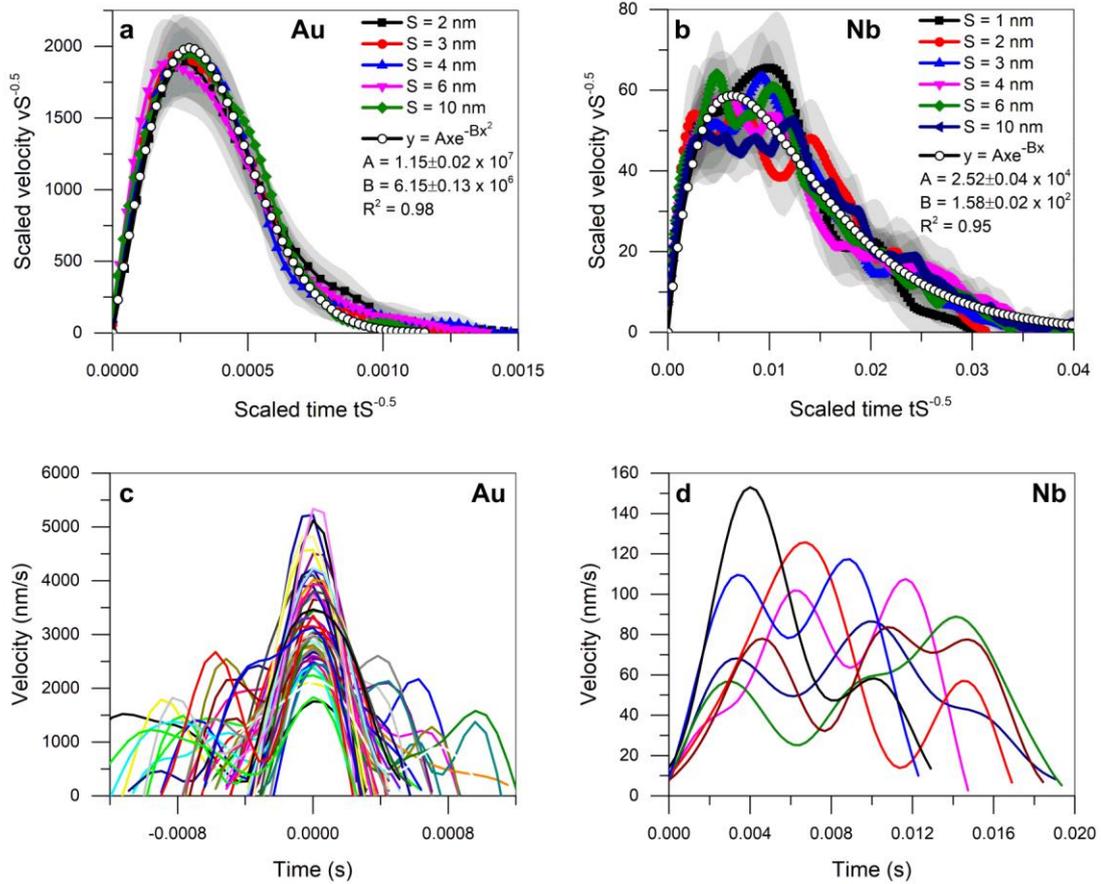

**Fig. 5.** Mean velocity profiles (averaged within narrow size bins) collapsed by scaling across different size bins for (a) gold and (b) niobium. Shaded areas represent 95% confidence intervals from the standard error of the mean for the velocity at a given time in each bin. Function in (a) is a scaling collapse function predicted by MFT with non-universal constants A and B determined by fitting. (c) shows velocity profiles for events in the 2 nm bin for the Au data with their peaks aligned at 0 s. (d) displays a selected subset of the velocity profiles within the 1 nm bin in the Nb data.

In order to demonstrate the general characteristics of the profiles in the two systems, Fig. 5c and Fig. 5d display velocity profiles from an individual size bin in the Au and Nb data, respectively. The Au profiles in Fig. 5c are displayed with their peaks aligned, showing the tendency for a single large peak with occasional smaller secondary peaks. The Nb profiles in Fig. 5d are instead displayed with their initial times aligned at t = 0, highlighting the trend towards a relaxation asymmetry, multiple peaks of similar magnitude, greater variability, and durations that are over 10 times longer. The profiles displayed in Fig. 5d are a selected subset of the profiles available in that size bin, as the greater variability in the Nb profiles causes a complete graph to become difficult to interpret.

While the used fitting functions produce reasonable fits to the data, the change in exponent required to better fit the Nb data raises the question of what results might be obtained if the exponent was one of the fitting parameters, i.e. using the fit equation $\langle v \rangle(t) = Ate^{-Bt^C}$ with $A$, $B$, and $C$ being fit parameters. Table 2 shows the resulting exponents and changes in goodness-of-fit for both the size-binned mean profiles shown in Fig. 5 and the size-scaled mean profiles shown in Fig. 6. The latter were obtained by scaling the velocity profiles for individual events in both time and velocity by a factor of $S^{-0.5}$ (where $S$ is the size of that particular event) and then averaging these rescaled profiles together to produce a single mean profile. Rescaling the profiles first and then averaging them together allows events of any size within the scaling regime to contribute to the mean profile. This contrasts with the size-binned profiles, where for a given mean profile only events within a narrow size range are selected to be averaged together directly, and then the mean profile itself is rescaled based on the central value of the size bin.

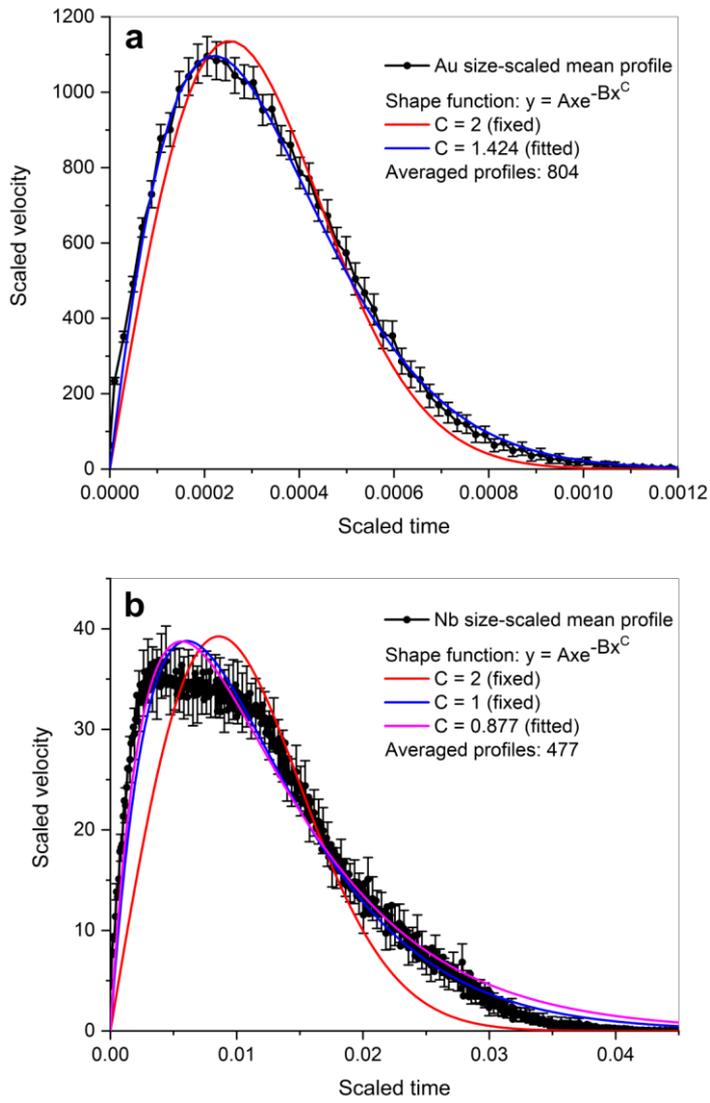

**Fig. 6.** Size-scaled mean velocity profiles for (a) Au and (b) Nb, with error bars representing the 95% confidence interval from the standard error of the mean, and shape function fits with various values of *C* (see Eq. 5 and Table 2).

| Material | Mean profile type | Exponent type | Exponent value | Goodness-of-fit value ($R^2$) |
|---|---|---|---|---|
| Au | Size-binned | Fixed | 2 | 0.980 |
| Au | Size-binned | Fitted | 1.662 | 0.986 |
| Au | Size-scaled | Fixed | 2 | 0.987 |
| Au | Size-scaled | Fitted | 1.424 | 0.996 |
| Nb | Size-binned | Fixed | 2 | 0.836 |
| Nb | Size-binned | Fixed | 1 | 0.946 |
| Nb | Size-binned | Fitted | 0.887 | 0.949 |
| Nb | Size-scaled | Fixed | 2 | 0.872 |
| Nb | Size-scaled | Fixed | 1 | 0.975 |
| Nb | Size-scaled | Fitted | 0.877 | 0.978 |
| Material | Mean profile type | Filtering method | Fitted relaxation-only exponent | Goodness-of-fit value ($R^2$) |
| Au | Size-scaled | FIR filtering | 1.997 | 0.999 |
| Nb | Size-scaled | FIR filtering | 1.404 | 0.998 |
| Au | Size-scaled | Wiener filtering | 1.508 | 0.999 |
| Nb | Size-scaled | Wiener filtering | 0.944 | 0.997 |

Table 2. Relaxation exponents and goodness-of-fit parameters for mean velocity profiles, including fits to only the velocity relaxation and a comparison of the FIR filtering method with an alternate filtering method (Wiener filtering).

For the size-binned profile collapses, the slight variation between the collapsed mean profiles for different bins limits how much the goodness-of-fit can be increased by allowing the relaxation exponent to be a fitting parameter. For the size-scaled mean profiles, only one profile is produced, and a much larger number of individual profiles contribute to the overall mean profile, reducing random variance and allowing a much tighter fit, especially for the Au data. For the Nb data, the perfection of the fit is limited by the lack of events with a scaled duration below a certain value, producing the noticeable "plateau" in the size-scaled mean profile after the initial peak but before the onset of the exponential relaxation, which cannot be reproduced by the fitting function. To eliminate any possible effect of this plateau on the observed difference in relaxation behavior, Table 2 also contains data from fits to only the velocity relaxation portion of the size-scaled mean profiles, starting at the peak velocity value for the Au data and at the end of the velocity plateau for the Nb data. For this fit, the time data is shifted to $t = 0$ at that point, and an alternate relaxation-only fitting function $\langle v \rangle(t) = Ae^{-Bt^D}$ is used to fit the relaxation data. It should be noted that the relaxation exponent $D$ found for this function will generally be different than the relaxation exponent $C$ found by fitting to the full profile, since the $t$-values are shifted relative to each other. Results are also presented for such a fit obtained by using an alternate displacement-filtering method (Wiener filtering) on the same raw nanoindenter test data; it can be seen that the precise value of the decay exponent is sensitive to details of the analysis method and that best fitting results are generally obtained for non-trivial exponents. Irrespective of the precise value of the exponent, the relatively slower nature of the Nb velocity relaxation is maintained in all cases.

To further demonstrate the extended velocity relaxation in the Nb crystal, Fig. 7 displays the probability distribution of the fraction of each individual velocity profile at which the peak velocity is reached. This quantity allows the determination of what fraction of the total event duration is taken up by the velocity relaxation from its peak value. It is found that the peak velocity for Au is often reached at a larger fraction of the duration (sample mean = 0.541, sample skewness = -0.209). For the case of Nb, the peak value is often reached earlier in the overall profile (sample mean = 0.495, sample skewness = 0.158), which demonstrates that the velocity relaxation in Nb is typically extended over a larger fraction of the duration of an avalanche, thereby supporting the interpretation of Fig. 5 as indicating slower velocity relaxation in the Nb crystal.

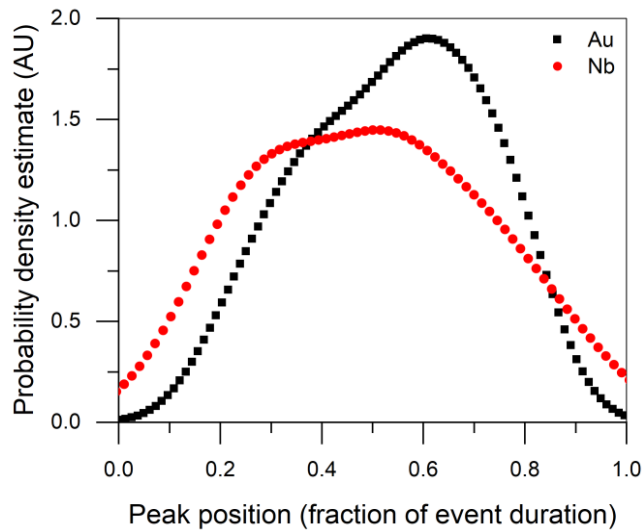

**Fig. 7.** Kernel smoothing function estimate of the probability distribution of relative peak position within event profiles for all events in the size bins specified in Fig. 3.

*3.6 Duration-binned mean velocity profile shapes*

In addition to examining size-binned mean profiles, we also consider duration-binned mean velocity profiles for both the Au and Nb data. To obtain the normalized mean-velocity profile shapes for events of similar duration, first a subset of events is chosen by event size (ranging from the smallest to largest sizes used for any of the size-binned profiles in that material, see Fig. 3) to ensure that all events are within the size-scaling regime. The events in that subset are binned by duration and normalized in duration and peak velocity. While the profiles within a given bin do not necessarily have the same number of points, they are very similar due to having identical DARs and similar durations. Points within the profiles are grouped using a second set of bins, with the number of grouping bins based on the average number of data points for profiles in that duration bin. Within these groups, velocity values are averaged to produce the mean value and 95% confidence interval of the mean for the velocity at that point, with the associated time value being the midpoint of the grouping bin. These mean profiles are then rescaled once more so that the time range spans from 0 to 1 and the highest mean velocity value is 1.

In the Au data, the profiles for the shortest-duration events are parabolic, with the profile shapes flattening and/or becoming skewed to the right as the duration increases. The latter tendency is consistent with the result (shown in Fig. 7) of the Au events having an increased probability of reaching their peak velocity value past the midpoint of the event duration. While inconsistent with the MFT predictions initially used here, asymmetry of duration-binned velocity profiles for avalanche events has also been reported in other theoretical and experimental work [39-43], with the explanations for said asymmetry varying depending on the nature of the system. The mean profile shapes for the Nb data are much more erratic due to the greater duration and variability within individual events. The mean profiles increase from zero to nearly the peak velocity rapidly, and then spends the majority of the event duration oscillating near that value (with a slight decreasing trend overall) until the event(s) finally terminate and the mean velocity rapidly decreases back to zero. Clearly, even with the more-symmetrical shape displayed in the duration-binned Nb profile shapes versus the size-binned Nb profile shapes, the Nb events spend proportionately much more time near their peak velocity (i.e., take proportionately much longer to relax back to zero velocity relative to the overall event duration) than the Au events.

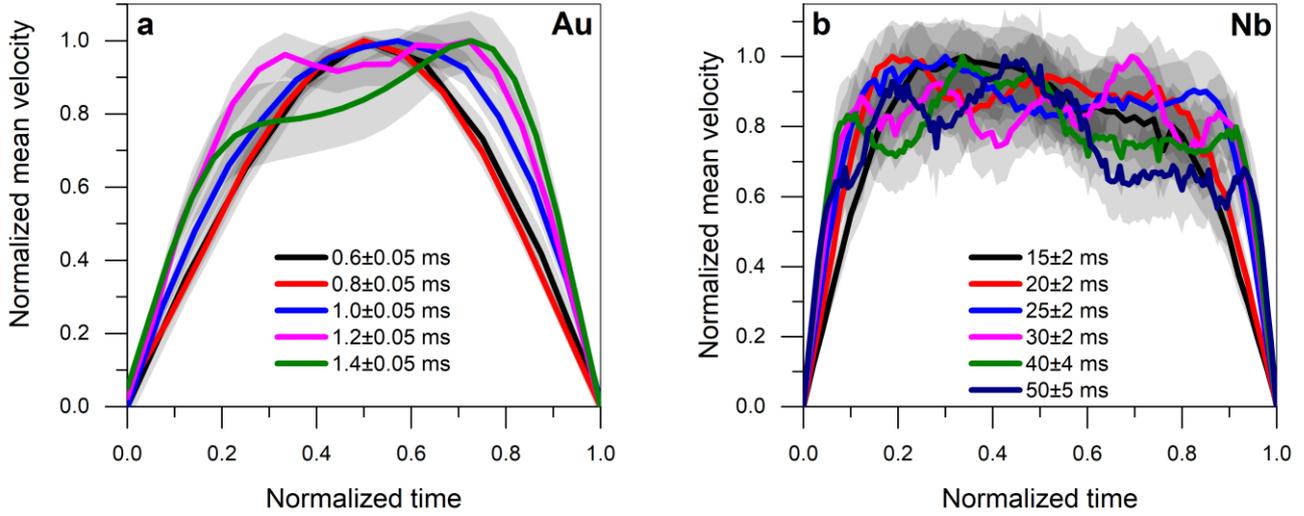

**Fig. 8.** Duration-binned and normalized mean velocity profile shapes for events in Au and Nb. Shaded areas represent 95% confidence intervals from the standard error of the mean for the normalized velocity at a given time in each bin.

*3.7 Origins of the differences in velocity relaxation behavior*

    The extended velocity relaxation in the BCC Nb relative to the FCC Au can be of two main origins, with either the instrument or the defect dynamics of the crystals themselves being the relaxation-limiting factor. We will first consider some critical aspects of the instrumentation. It is known that the nanoindentation system behaves as a damped harmonic oscillator [18], which for these tests was found to have a physical resonance frequency of 154 Hz. This corresponds to a period of approximately 6.5 ms. The Au events all have durations of less than 5 ms, with 90% having durations below 2 ms. They are thus unlikely to have been significantly affected by this resonance, having occurred on a substantially shorter time scale. However, the Nb events have durations ranging from 6 to 187 ms (12 to 134 ms if only considering events in the size-scaling regime). This puts a significant fraction of the events in a similar duration range as the natural impulse response of the indenter system, which begins reacting significantly after one oscillation period (i.e., ~6.5 ms in this case) and damps out to a small amplitude in 4-5 oscillation periods (i.e., ~30 ms) [18]. Thus, the indenter impulse response might potentially be expected to have an effect on events in this duration range. However, if this were to be the case, it would be expected that an oscillating force signal due to the impulse response would be observed during the

affected events, which is not the case. Indeed, the accelerations that occur during the Nb slip events are remarkably low, to the point that the indenter tip is effectively in quasi-static equilibrium with the sample even during "dynamic" slip. This means the material must be controlling the deformation behavior, not the machine. Similarly, if oscillations of increased force resulting from the device's resonant impulse response were to affect the observed velocity profile, one would expect local maximums in the velocity profile to occur consistently at similar time intervals from the beginning of the event, which is not observed (see e.g. Fig. 5d). There are some local maximums in certain velocity profiles that occur at identical times, but in other velocity profiles we find local *minimums* occurring at the same time relative to the beginning of the event, or sections that are neither a local minimum or maximum. Additionally, events that are significantly longer in duration than the impulse response of the indenter would be expected to be less affected by the impulse response due to its decay over time; if the size-scaled mean profile is restricted to events of size 5 to 12 nm (resulting in a minimum duration of 33 ms), the extended velocity relaxation is still maintained. The lack of resonance effects was further confirmed via wavelet analysis of both the displacement-time and velocity-time data. The dominant frequency components present in the velocity-time data for individual slip events were widely scattered for both the Au and Nb data sets, and very few events had dominant components near the 154 Hz resonance frequency of the indenter, with 98% of the Au events having dominant frequencies above 200 Hz and 99% of the Nb events having dominant frequencies below 100 Hz. The displacement-time data for individual slip events was always dominated by very low-frequency components, due to the displacement jumps being significantly linear. As previously mentioned in Section 3.4, the extended velocity relaxation for Nb slip events was also found to be maintained regardless of machine control mode (closed-loop displacement control, closed-loop force control, or open-loop force control). We therefore conclude that the pronounced difference in the velocity relaxation between Au and Nb is unlikely to be a machine effect, and turn to a possible microstructural explanation.

Underlying the avalanche profiles examined in Figs. 2, 5, 6, and 8 is a collective dislocation rearrangement that cannot be traced in experiments, but that in the case of microcrystals has been shown to be dominated by spiral dislocation sources [44, 45]. Due to the higher lattice friction and low screw-dislocation mobility in BCC crystals, we adopt the view that the clear difference in dislocation avalanche velocity relaxation is governed by the screw-component of a mixed character spiral-arm source. This can be substantiated on the basis of recent in-situ transmission electron microscopy (TEM) results by Caillard [46], where spiral-arm sources were tracked during deformation of pure BCC iron. It was found that the edge-component of a mixed dislocation arm rapidly leaves the crystal, leaving a slowly propagating straight screw-dislocation behind. Thus, for each 180° rotation of the spiral-arm source, the highly mobile edge-components will initially (i.e. at the beginning of the 180° rotation) produce fast crystallographic slip, followed by a gradual velocity decay that is dictated by the motion of the slow screw-components. Clearly, several such mixed spiral-arm sources are required to generate the net displacement observed as a result of the occurring avalanche, and stochastic activation of new sources during an avalanche in the BCC crystals could lead to the oscillating velocity profiles observed (see Fig. 5d) due to the introduction of new fast edge-components. Since the velocity of edge- and screw-dislocations in FCC crystals are practically identical [47], the averaged avalanche profile would maintain a more symmetric shape without strong fluctuations during the velocity relaxation, as is observed. We therefore conclude that the difference in velocity-relaxation behavior between Au and Nb is a manifestation of a material-specific avalanche dynamics that is not seen in the very similar truncated power-law statistics of the corresponding slip-sizes.

## 4. Concluding remarks

In summary, we trace the time-resolved dislocation avalanche velocity profiles in Au and Nb microcrystals. Both crystal types exhibit scale-invariant slip-size distributions with scaling exponents for the size statistics and parts of the dynamics that appear to agree with mean-field theory. However, the dynamic shape scaling function shown here reveals fundamentally different avalanche-velocity relaxation, where the slower velocity decay for Nb is rationalized by screw-limited avalanche dynamics. While the mean-field theory predictions fit the collapsed average avalanche-shape well for the Au data, a detailed investigation of the average velocity shape functions reveals that the best fitting results are obtained with exponents different than the tested theoretical functions for both data sets. In particular, the best-fit exponents describing the velocity relaxation are significantly lower than theory would predict, especially for the Nb data. Irrespective of the precise form of the shape function, the velocity relaxation remains substantially slower for Nb in comparison to Au independent of experimental details, e.g. the loading and steering modes. We therefore conclude that this study reveals details of the slip dynamics that distinctly depend on the material specifics, suggesting non-universal behavior, which is also seen when comparing different FCC crystals amongst each other [48]. We anticipate that the here used method can be useful to gain deeper insights into collective dislocation dynamics across different materials, and in particular of BCC crystals, where local core effects determine the mobility of dislocations.

**Acknowledgements**

This research was carried out in part in the Frederick Seitz Materials Research Laboratory Central Research Facilities, University of Illinois. G.S. and R.M. especially thank Kathy Walsh for experimental support with the Hysitron TriboIndenter. R.M. would like to thank P.M. Derlet for fruitful discussions, and is grateful for financial support by the NSF CAREER program (grant NSF DMR 1654065), and for start-up funds provided by the Department of Materials Science and Engineering at UIUC.